\documentclass{sig-alternate}
\usepackage{mathptmx} % This is Times font

\usepackage{fancyhdr}
\usepackage[normalem]{ulem}
\usepackage[hyphens]{url}
\usepackage[sort,nocompress]{cite}
\usepackage[final]{microtype}
\usepackage[keeplastbox]{flushend}
\usepackage[bookmarks=true,breaklinks=true,letterpaper=true,colorlinks,linkcolor=black,citecolor=blue,urlcolor=black]{hyperref}
\usepackage{url}
\usepackage{threeparttable}
\usepackage{tikz}
\usepackage{amsmath}
\usepackage{booktabs}
\usepackage{subfig}
\usepackage{comment}
\usepackage{pifont}
\usepackage{soul}
\usepackage{multirow}

\newcommand{\squishlist}{
   \begin{list}{$\bullet$}
    { 
    \setlength{\itemsep}{0pt}      \setlength{\parsep}{0pt}
      \setlength{\topsep}{3pt}       \setlength{\partopsep}{0pt}
      \setlength{\listparindent}{-2pt}
      \setlength{\itemindent}{-5pt}
      \setlength{\leftmargin}{1em} \setlength{\labelwidth}{0em}
      \setlength{\labelsep}{0.5em} } }

\newcommand{\squishend}{
    \end{list}  }

\fancypagestyle{firstpage}{
  \fancyhf{}
  
  \fancyhead[C]{\vspace{15pt}\normalsize{}} 
  \fancyfoot[C]{\thepage}
}

\title{Towards Efficient Architecture and Algorithms for Sensor Fusion} 
\begin{document}

\author{%
\and
Zhendong Wang\\
  \affaddr{University of Texas at Dallas}\\
  \affaddr{Richardson, Texas, USA}\\
  %\email{Zhendong.Wang@utdallas.edu}
\and
Xiaoming Zeng\\
  \affaddr{University of Texas at Dallas}\\
  \affaddr{Richardson, Texas, USA}\\
  %\email{Xiaoming.Zeng@utdallas.edu}
\and
Shuaiwen Leon Song\\
  \affaddr{University of Sydney, Australia}\\
  \affaddr{Sydney, Australia}\\
  %\email{tax6@pitt.edu}
\and
Yang Hu\\
  \affaddr{University of Texas at Dallas}\\
  \affaddr{Richardson, Texas, USA}\\
  %\email{hu\_yang@tsinghua.edu.cn}
}

\maketitle
\thispagestyle{firstpage}
\pagestyle{plain}

%%%%%% -- PAPER CONTENT STARTS-- %%%%%%%%

\begin{abstract}
The safety of an automated vehicle hinges crucially upon the accuracy of perception and decision-making latency. Under these stringent requirements, future automated cars are usually equipped with multi-modal sensors such as cameras and LiDARs. The sensor fusion is adopted to provide a confident context of driving scenarios for better decision-making. A promising sensor fusion technique is middle fusion that combines the feature representations from intermediate layers that belong to different sensing modalities. However, achieving both the accuracy and latency efficiency is challenging for middle fusion, which is critical for driving automation applications. 

We present $A^{3}$Fusion, a software-hardware system specialized for an \textit{adaptive}, \textit{agile}, and \textit{aligned} fusion in driving automation. $A^{3}$Fusion achieves a high efficiency for the middle fusion of multiple CNN-based modalities by proposing an adaptive multi-modal learning network architecture and a latency-aware, agile network architecture optimization algorithm that enhances semantic segmentation accuracy while taking the inference latency as a key trade-off. In addition, $A^{3}$Fusion proposes a FPGA-based accelerator that captures unique data flow patterns of our middle fusion algorithm while reducing the overall compute overheads. We enable these contributions by co-designing the neural network, algorithm, and the accelerator architecture.
\end{abstract}
\vspace{-3pt}
\section{Motivation}
\vspace{-3pt}
Autonomous vehicles are typically equipped with multiple different sensors, and thus, multi-sensing modalities can be fused to exploit their complementary properties to achieve robust and accurate scene understanding, such as semantic segmentation \cite{feng2020deep}. Generally, fusion can be achieved at the input level, decision level, or intermediately (i.e., middle fusion).    
Recently, several methods adopting deep learning have been proposed to fuse multi-modal sensors to implement semantic segmentation for
autonomous driving \cite{sun2020real, zeng2022efficient}. Even so, it is still an open issue how to achieve the optimal way to effectively fuse the multiple modalities, especially via the complicated middle fusion due to the fact that the middle fusion is highly challenging. 
Therefore, our work focuses on RGB-Depth information fusion for semantic segmentation in autonomous driving and targets at an efficient architecture and algorithms to achieve middle fusion to enhance scene perception.
 
Regarding the middle fusion, the intermediate level features from different modalities typically possess misaligned spatial dimensions (e.g., different channels), making it infeasible to simply concatenate or element-wise add the features. Therefore, \textbf{effective solutions are required to address the spatial-feature-misalignment issue}.

Secondly, middle fusion provides rich intermediate features and offers a large range of choices to combine sensing modalities, and we cannot easily conclude one fusion implementation is better than the others. Therefore, \textbf{specific strategies are needed to implement middle fusion to effectively enhance segmentation performance}.   

Thirdly, most works conducting fusions only consider segmentation accuracy and robustness. However, autonomous driving poses importance on latency \cite{Wang2020Understanding, Wang2020enabling}. Therefore, \textbf{it is necessary to take latency performance into consideration in middle fusion besides the accuracy and robustness}.
Besides, to implement the fusion, the multiple modalities (e.g., two DNNs) have to co-execute simultaneously \cite{Bateni2020co}, and the introduced fusion unavoidably incurs extra computation cost and data dependency between the two modalities, which leads to significant inefficiency on existing GPU platforms or accelerators and calls for a more customized hardware design to amortize computation efficiency and data synchronization issues \cite{Baek2020multi, Maor2019fpga}.

\section{Main Artifacts}
We present $A^{3}$Fusion, a software-hardware system specialized for an \textit{adaptive}, \textit{agile}, and \textit{aligned} fusion in driving automation. $A^{3}$Fusion achieves a high efficiency for the middle fusion of multiple CNN-based modalities by proposing an adaptive multi-modal learning network architecture and a latency-aware, agile network architecture optimization algorithm that enhances semantic segmentation accuracy while taking the inference latency as a key trade-off. In addition, $A^{3}$Fusion proposes a FPGA-based accelerator that captures unique data flow patterns of our middle fusion algorithm while reducing the overall compute overheads. We enable these contributions by co-designing the neural network, algorithm, and the accelerator architecture.

\subsection{fuseLinks for Accuracy in Middle Fusion}\label{introfuseLink}
To address the spatial-feature-misalignment issue in intermediate-level features as well as promote the information integration across different channels of the fused features, we propose a \textit{fuseLink} to implement middle fusion, which includes a convolution filter (i.e., \textit{fuseFilter}) in each connection. For two layers with misaligned spatial features, the fuesLink adopts the dimension of which is $c_{input}$ * $c_{output}$ * w * h, to adapt the channel number of outputted feature of active fusion layer to exactly match the channel number of inputted feature of passive fusion layer. Also, as the network goes deeper, the fuseLink provides an opportunity to fuse the low-level information to the high-level features instead of merely summarizing the same level information, benefiting the context-critical semantic segmentation. Besides for two layers with matched spatial features, the fuseFilter can promote the information integration across different channels of the spatial features to improve the segmentation performance.  
\subsection{How To Effectively Deploy fuseLinks}\label{deploypolicy}
Even though fuseLinks can be applied to implement the middle fusion smoothly, 
it is non-trivial to decide the "optimal" way to deploy the fuseLinks. By comprehensive characterizations, we explore the deployment principles from: the fusion direction, the fusion distance and the fusion number as well as fusion location. Specifically, bidirectional fuseLinks outperforms unidirectional one and the more fuseLinks are applied, the more accuracy benefit are obtained. Meanwhile, the fuseLink with distance = 1 can bring the largest benefit compared to the fuseLinks with other distance. Besides, the fuseLinks connecting the low-level blocks typically outperforms the fuseLinks on the high-level blocks. By following the principles, fuseLinks deployment can be adapted to different scenarios and be wisely adjusted optimally to improve the segmentation accuracy.

\subsection{Latency-aware fuseLinks Deployment}\label{latencyfusion}
\begin{figure}[t]
\centering
            \includegraphics[width=1\linewidth]{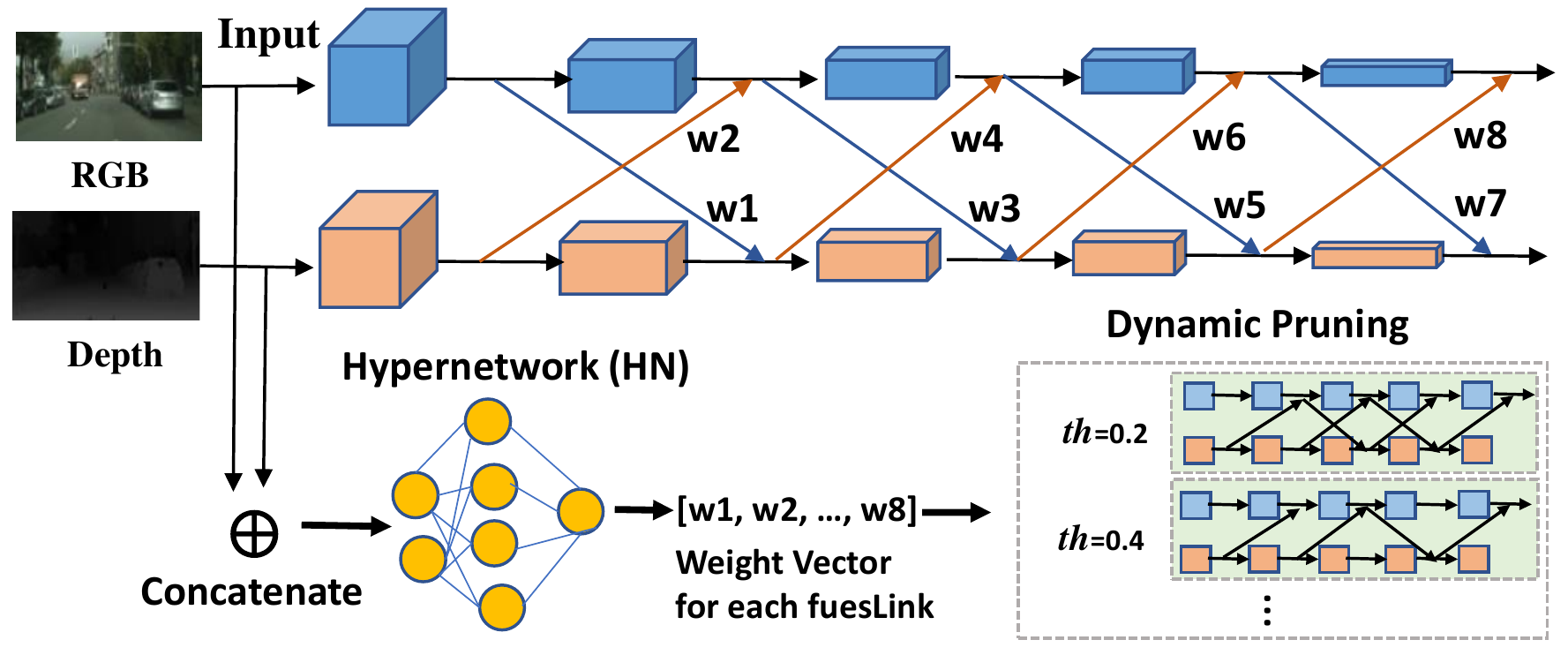}
            \vspace{0mm}
            %\captionsetup{font={normal}}
            \caption{Latency-aware middle fusion.}\label{latencyincludednet}
            %\vspace{-5mm}
\end{figure}

As fuseLinks unavoidably incur extra computation cost and data dependency between RGB and depth, which may cause extra latency compared to the individual RGB or depth network, this is unacceptable to autonomous driving safety. Therefore, we propose a latency-aware fuseLink pruning approach to prune the fuseLinks that contributes little to the ultimate accuracy to reduce the computation cost and enhance the latency performance of the segmentation.
As Fig. \ref{latencyincludednet} shows, we construct a hypernetwork (HN) that can decide the importance of each link to the ultimate accuracy depending on the input data. HN is composed of one convolution layer, average Pooling layer and flattening layer as well as one FC layer and  
%HN is trained with the RGB and depth networks together after the fuseLinks are deployed, and  will generate a weight, $w_{l}$, for each fuseLink after being well trained. 
$w_{l}$ can change dynamically depending on the specific input in inference, and we set a threshold, $th$, which can be adjusted to wisely decide which fuseLinks can be pruned for a given input to trade-off the accuracy with latency of the segmentation. 
In practice, autonomous driving may drive in different scenarios, such as the rural road, urban intersection, and each scenario imposes different requirements on the accuracy and latency of vehicle's perception function and reaction. By adapting threshold, we can trade off the latency and accuracy such that the vehicle can drive safely and efficiently in different scenarios. 

\section{Fuse-Multitasking Accelerator Design} \label{fusemt}
As the fuseLinks can be adapted to specific fusion scenarios and support dynamic pruning, leading to the hardware underutilization. Meanwhile, the remained fuseLinks imposes data dependencies between the two fused networks, causing computation inefficiency when running in traditional DNNs accelerators. 
Therefore, we develop an aligned accelerator architecture on FPGA, to thoroughly explore the fuseLinks' contribution to the segmentation accuracy while ease the extra computation and latency cost.

\begin{figure}[t]
\centering
            \includegraphics[width=\linewidth]{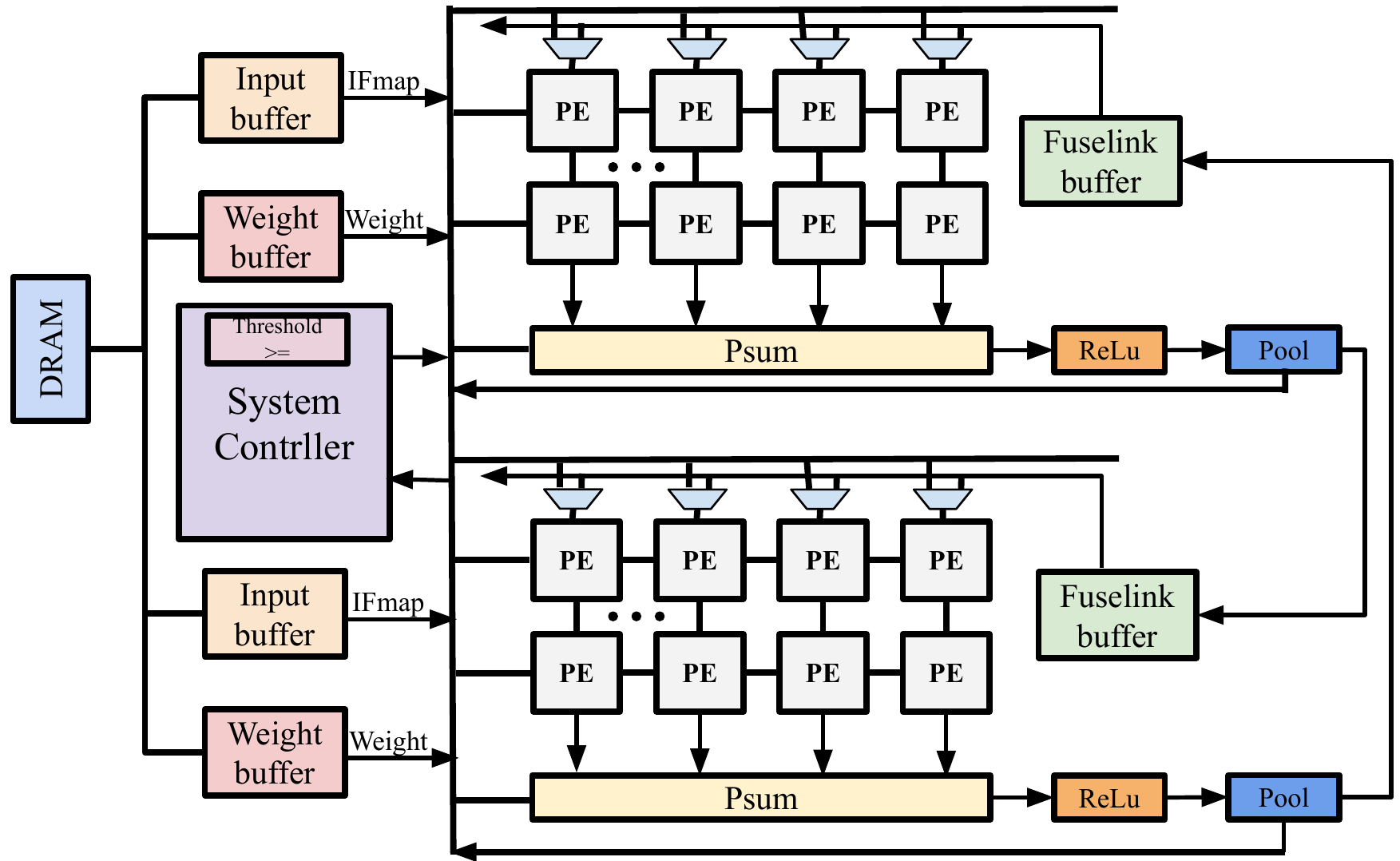}
            \caption{Specific architecture of Fuse-Multitasking.}\label{Overview2}
\end{figure}

Fig. \ref{Overview2} shows our customized aligned accelerator architecture. Basically, we split a large PE array into half and each half forms an individual processing system which is configured with its own buffers and controlling logic. Even though the two "independent" parts can execute in parallel to implement RGB and depth network on their own hardware system, executing the RGB and depth networks on their own. Meanwhile, the specialized fuseLink buffer effectively connect the two PE part and shares the partial sum results during the fusion such that the RGB and depth network can almost execute in an aligned mode. 

\section{Key Results and Contributions}
\noindent\textbf{Accelerator setup} All hardware modules of $A^{3}$ Fusion are implemented in Verilog and synthesized on a XC7VX690T-2FFG1761C Xilinx FPGA. The frequency is 80MHz. Each PE performs one MAC operation with total 5 hard macro on-chip DSP IPs. In addition, we adopt 32-bit floating-point DSP operation in order to maintain the calculation accuracy. The performance results including inference latency, resources utilization and power consumption are evaluated via Vivado 2018.3 version. 

\noindent\textbf{Segmentation accuracy analysis and its impact on vehicle safety}
Firstly, by extensively evaluate how our fuseLinks improve the segmentation accuracy and how the different deployment principles impact the accuracy specifically, we observe that  
the bidirectional fuseLink with distance = 2 outperforms all the other fuseLink deployment, and reaches 37.66\%. Compared to the baseline FuseNet \cite{Hazirbas2016fusenet}, it improves 2.19\%, indicating significant risk reduction in vehicle driving. More details can be found in full paper. 

%\begin{figure}[t]
%\centering
%            \includegraphics[width=\linewidth]{./picture/acceveluatesum.pdf}
%            \caption{Specific architecture of Fuse-Multitasking.}\label{Overview2}
%\end{figure}
\noindent\textbf{$A^{3}$Fusion Accelerator in terms of latency, resource utilization and power}
%To evaluate the performance of our accelerator in terms of latency, resources utilization and power, we set two configurations: baseline and $A^{3}$ Fusion. Baseline indicates the implementation of fuseLinks and the two RGB and depth networks on one complete PE array which is configured with 16x16 and no specialized fuseLink buffer and MUX. In comparison, $A^{3}$ Fusion indicates the implementation on our customized aligned architecture, which adopts two independent PE arrays with 8x16 and configures fuseLink buffer to cache psums and MUX to dynamically prune links to speed up the fusion.
$A^{3}$ Fusion hardware architecture can reduce the segmentation latency up to 17.97\% compared to the baseline. 
%The main constraint is that the two VGG-16 branches of the main network hold the most part of MACs operations. 
If the dynamic pruning techniques applied, the more fuseLink are deployed, the more benefit our $A^{3}$ Fusion architecture can bring.
Concerning the resources utilization, 20.84\% LUT and 27.43\% FF overhead are incurred due to that one large PE arracy (i.e., baselien) is split into half and extra controlling logic is introduced for the fuseLink.
Also, as each half PE array is configured with sperate fuseLink buffer for effective communication,  73.78\% storage overhead is incurred.
More details can be found from full paper.

\section{Conclusion}
The sensor fusion is playing a vital role in future driving automation system as it guarantees a more accurate object detection and semantic segmentation. However, trading-off the accuracy and latency of multi-modal fusion is still unexplored. We propose an optimized multi-model fusion network in this work and comprehensively characterize the design space. The key to our approach is to co-design the neural network, algorithm, and hardware accelerator to achieve the optimized trade-off between accuracy and latency in sensor fusion. Our work also promises to extend to a more generalized multi-modal fusion scenario.


\begin{thebibliography}{00}

\bibitem{feng2020deep} Feng, D., Haase-Schütz, C., Rosenbaum, L., Hertlein, H., Glaeser, C., Timm, F., ... \& Dietmayer, K. (2020). Deep multi-modal object detection and semantic segmentation for autonomous driving: Datasets, methods, and challenges. IEEE Transactions on Intelligent Transportation Systems, 22(3), 1341-1360.

\bibitem{sun2020real} Sun, L., Yang, K., Hu, X., Hu, W., \& Wang, K. (2020). Real-time fusion network for RGB-D semantic segmentation incorporating unexpected obstacle detection for road-driving images. IEEE Robotics and Automation Letters, 5(4), 5558-5565.
%\bibitem{chen2019progressive} Chen, Z., Zhang, J., \& Tao, D. (2019). Progressive lidar adaptation for road detection. IEEE/CAA Journal of Automatica Sinica, 6(3), 693-702.

\bibitem{zeng2022efficient} Zeng, X., Wang, Z., \& Hu, Y. (2022). Enabling Efficient Deep Convolutional Neural Network-based Sensor Fusion for Autonomous Driving. arXiv preprint arXiv:2202.11231.

%\bibitem{Wang2022Towards} Wang, Z., \& Hu, Y. (2022). Towards a High-performance and Secure Memory System and Architecture for Emerging Applications. arXiv preprint arXiv:2205.04002.

\bibitem{Wang2020Understanding} Wang, Z., Wang, Z., Liu, C., \& Hu, Y. (2020). Understanding and tackling the hidden memory latency for edge-based heterogeneous platform. In 3rd USENIX Workshop on Hot Topics in Edge Computing (HotEdge 20).

\bibitem{Wang2020enabling} Wang, Z., Jiang, Z., Wang, Z., Tang, X., Liu, C., Yin, S., \& Hu, Y. (2020). Enabling Latency-Aware Data Initialization for Integrated CPU/GPU Heterogeneous Platform. IEEE Transactions on Computer-Aided Design of Integrated Circuits and Systems, 39(11), 3433-3444.

\bibitem{Bateni2020co} Bateni, S., Wang, Z., Zhu, Y., Hu, Y., \& Liu, C. (2020, April). Co-optimizing performance and memory footprint via integrated cpu/gpu memory management, an implementation on autonomous driving platform. In 2020 IEEE Real-Time and Embedded Technology and Applications Symposium (RTAS) (pp. 310-323). IEEE.

\bibitem{Baek2020multi} Baek, E., Kwon, D., \& Kim, J. (2020, May). A multi-neural network acceleration architecture. In 2020 ACM/IEEE 47th Annual International Symposium on Computer Architecture (ISCA) (pp. 940-953). IEEE.

\bibitem{Maor2019fpga} Maor, G., Zeng, X., Wang, Z., \& Hu, Y. (2019, November). An FPGA implementation of stochastic computing-based LSTM. In 2019 IEEE 37th International Conference on Computer Design (ICCD) (pp. 38-46). IEEE.

\bibitem{Hazirbas2016fusenet} Hazirbas, C., Ma, L., Domokos, C., \& Cremers, D. (2016, November). Fusenet: Incorporating depth into semantic segmentation via fusion-based cnn architecture. In Asian conference on computer vision (pp. 213-228). Springer, Cham.

\end{thebibliography}
\end{document}